\documentclass[aps,pre,reprint,superscriptaddress,showkeys,amsmath,amssymb,twocolumn,nofootinbib]{revtex4-2}
\usepackage{orcidlink}
\usepackage{microtype}
\usepackage[separate-uncertainty=true, multi-part-units=single]{siunitx}
\usepackage[version=4]{mhchem}
\usepackage{graphicx}
\graphicspath{{figs/}}

\usepackage{hyperref}
\hypersetup{colorlinks, breaklinks, urlcolor=blue, linkcolor=blue, citecolor=blue, anchorcolor=blue}

\begin{document}
\title{Active Brownian motion in a single-relaxation viscoelastic fluid}
\author{Sanatan Halder~\orcidlink{0009-0002-2457-0449}}
\email{sanatanh@iitk.ac.in}
\affiliation{Department of Physics, Indian Institute of Technology Kanpur, Kanpur -- 208016, India}
\author{Manas Khan~\orcidlink{0000-0001-6446-3205}}
\email{mkhan@iitk.ac.in}
\affiliation{Department of Physics, Indian Institute of Technology Kanpur, Kanpur -- 208016, India}
\begin{abstract}
	
	Active Brownian particles (ABPs) in viscoelastic (VE) media exhibit fascinating dynamical phenomena set by self-propulsion, thermal fluctuations, and fluid viscoelasticity. We extend our model, in which the Brownian dynamics within a slowly diffusing harmonic well emulates that in a single-relaxation VE fluid, to study active Brownian motion in such media. Consequently, the resultant dynamics is governed by the interplay of the characteristic timescales of the systems: the crossover and equilibration times of the VE fluid, $\tau_k$ and $\lambda$, respectively, and the persistence time of the ABP, $\tau_{\mathrm{R}}$. Following analytical predictions and simulations, we study two practically relevant regimes where the dynamics is dominated by the persistence of active motion and the elastic confinement of the VE fluid, with a phoretically active Pt-coated Janus colloid in a dynamic optical trap, and show quantitative agreement with the simulations. This approach provides a VE environment with tunable VE properties that remain unaffected by the strength of self-propulsion, allowing us to systematically investigate active Brownian motion in VE media in ways that are not otherwise possible with physical VE fluids.

\end{abstract}
\maketitle

The dynamics of active particles in viscoelastic (VE) media produce phenomena absent in Newtonian solvents, arising from the interplay of fluid memory with the self-propulsion and thermal fluctuations of the particle~\cite{elgetiPhysicsMicroswimmersSingle2015, bechingerActiveParticlesComplex2016, zottlEmergentBehaviorActive2016, liMicroswimmingViscoelasticFluids2021}. 
This interplay is widespread in biology, where bacteria swim through polymeric solutions~\cite{pattesonRunningTumblingColi2015, martinezFlagellatedBacterialMotility2014, magariyamaMathematicalExplanationIncrease2002}, infiltrate VE biofilm matrices~\cite{houryBacterialSwimmersThat2012}, spermatozoa traverse VE reproductive fluids~\cite{fauciBIOFLUIDMECHANICSREPRODUCTION2006, smithBendPropagationFlagella2009}, and pathogens penetrate mucus by locally altering its rheology~\cite{celliHelicobacterPyloriMoves2009, mirbagheriHelicobacterPyloriCouples2016}.
On the synthetic side, self-propelled Janus colloids in VE fluids give controlled experimental access to the coupling between activity and fluid memory~\cite{gomez-solanoDynamicsSelfPropelledJanus2016, lozanoRunandtumblelikeMotionActive2018, saitoSelfpropelledMotionInducedcharge2025, narinderActiveColloidsGeometrical2021, narinderActiveParticlesGeometrically2019, narinderWorkFluctuationRelation2021, tanukuActiveParticlesTunable2026, saadDiffusiophoresisActiveColloids2019}.
Related work has examined how VE memory modifies active Brownian motion through generalized Langevin approaches~\cite{sprengerActiveBrownianMotion2022, quevedoActiveBrownianParticles2026} and simulations in polymer solutions and networks~\cite{duStudyActiveBrownian2019, yuanActivityCrowdingCoupling2019, kumarDynamicsSelfpropelledTracer2023, teranViscoelasticFluidResponse2010, jooAnomalousDiffusionActive2020}.
A single-relaxation VE fluid, however, provides the minimal realization of this interplay, where the dynamics of a self-propelled particle remain tractable.

For passive particles in single-relaxation VE fluids, such as wormlike micellar solutions, the dynamics are understood within the Jeffreys or Maxwell-Voigt (MV) model~\cite{catesStaticsDynamicsWormlike1990, hassanMicrorheologyWormlikeMicellar2005, vanzantenBrownianMotionSingle2000, grimmBrownianMotionMaxwell2011, raikherTheoryBrownianMotion2010, raikherBrownianMotionViscoelastic2013}.
A passive Brownian particle (PBP) in an MV fluid is described by the harmonically bound Brownian particle (HBBP) with long-time diffusion model, where the confining potential itself diffuses at the VE relaxation time $\lambda$~\cite{khanRandomWalksColloidal2014, halderRealizingMicrorheologicalResponse2026a}.
The resulting mean square displacement (MSD) crosses over from Brownian diffusion through an elastic plateau to a second diffusive regime beyond $\lambda$.
A self-propelled particle adds directional persistence to this picture, alongside the elastic confinement and VE relaxation already present.
How these three effects compete to shape the MSD, in particular its superdiffusive regime, plateau, and onset of long-time diffusion, remains less explored.

Here we show that an active Brownian particle (ABP) in a single-relaxation VE fluid is described by a harmonically bound active Brownian particle (HBABP) with long-time diffusion, extending the static-well HBABP framework~\cite{halder2025interplay, halderIdentifyingSignaturesResidual2026} to a diffusing harmonic well (HW)~\cite{khanRandomWalksColloidal2014, halderRealizingMicrorheologicalResponse2026a}.
Three competing timescales govern these dynamics: the equilibration time $\tau_k = \gamma_{\mathrm{s}}/k$, the persistence time $\tau_{\mathrm{R}} = 1/D_{\mathrm{R}}$, and the VE relaxation time $\lambda = \gamma/k$, where $\gamma_{\mathrm{s}}$ and $\gamma \equiv \gamma_{\mathrm{HW}}$ are the high- and low-frequency drags, the latter from the diffusing HW.
The ordering of these three timescales sets distinct dynamical regimes.
To access a subset of these regimes experimentally, we realize ABP motion in a configurable single-relaxation VE medium by placing a phoretically active Pt-coated silica Janus colloid~\cite{halderDynamicallyStableOptical2026} in an optical trap steered along a Brownian trajectory~\cite{halderRealizingMicrorheologicalResponse2026a}.
The laser power $P$ sets the trap stiffness $k$, and the imposed HW diffusivity $D_{\mathrm{HW}}$ sets the low-frequency friction $\gamma_{\mathrm{HW}} = k_{\mathrm{B}}T/D_{\mathrm{HW}}$ at temperature $T$, so both controls jointly fix $\lambda$.
Because the Janus colloid is suspended in a purely viscous solvent, with no polymeric or micellar network, the VE response arises entirely from the trap dynamics.
The absence of a network also leaves activity nothing to perturb, so the effective medium remains unaltered~\cite{halderDynamicallyStableOptical2026}.
Theory, simulation, and experiment then resolve the MSD signature of each accessible regime in quantitative agreement.

\begin{figure}[t]
    \centering
    \includegraphics[width=\linewidth]{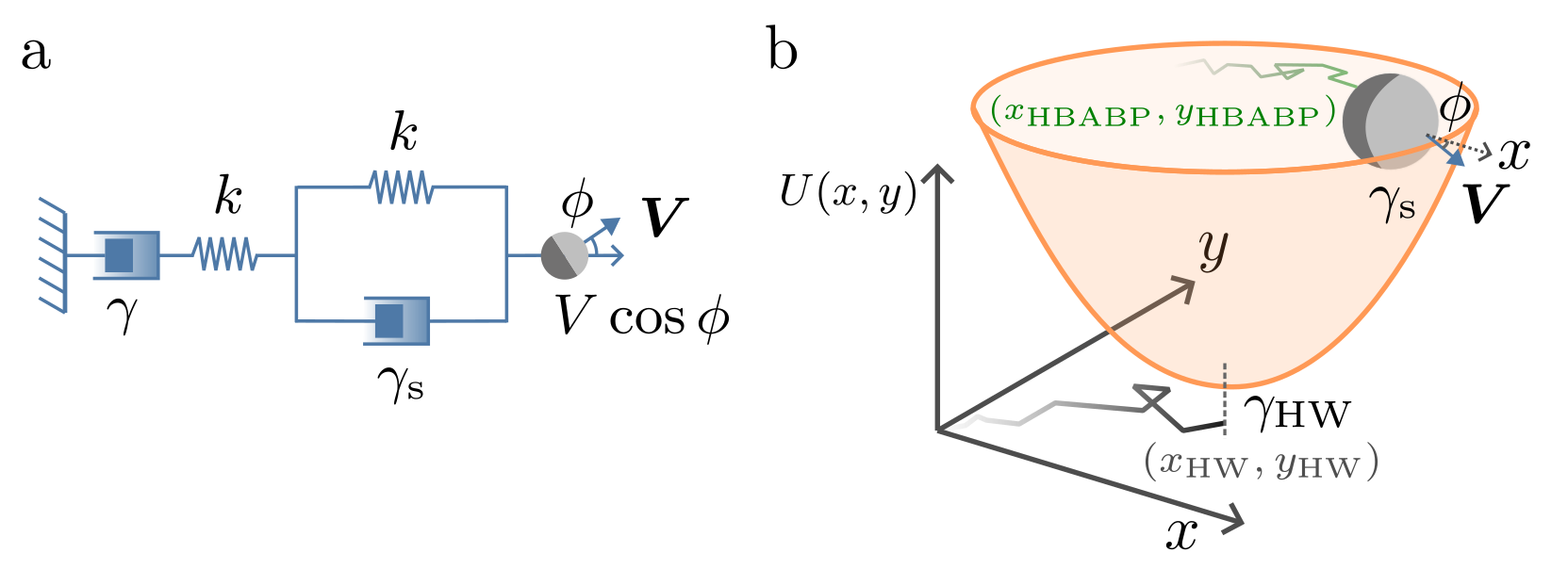}
    \caption{Rheological model of an active Brownian particle (ABP) in Maxwell-Voigt (MV) fluid and its stochastic representation as a harmonically bound active Brownian particle (HBABP) with long-time diffusion.
        (a)~A Voigt element, spring $k$ in parallel with dashpot $\gamma_{\mathrm{s}}$, is connected in series with a Maxwell element, spring $k$ in series with dashpot $\gamma$.
        Both springs share the same stiffness $k = 6\pi a G_{\mathrm{p}}$, which sets the elastic plateau modulus $G_{\mathrm{p}}$ for an ABP of Stokes radius $a$.
        Here $\gamma_{\mathrm{s}}$ and $\gamma$ represent the high- and low- frequency friction contributions, respectively.
        (b)~The ABP, shown by the green trajectory, moves in a harmonic well (HW) of stiffness $k$ and solvent friction $\gamma_{\mathrm{s}}$, while the HW center, shown by the gray trajectory, diffuses with friction $\gamma_{\mathrm{HW}}$ analogous to the Maxwell friction $\gamma$ in~(a).}%
    \label{fig:hbabp-ltd}%
\end{figure}

\paragraph*{Model.} An MV fluid connects a Voigt element, spring $k$ in parallel with dashpot $\gamma_{\mathrm{s}}$, in series with a Maxwell element, the same spring $k$ in series with dashpot $\gamma$ (Fig.~\ref{fig:hbabp-ltd}a).
The Voigt element confines the particle harmonically, while the Maxwell element adds slow diffusion of the HW center beyond the relaxation time $\lambda$.
An ABP in this model fluid therefore behaves as an HBABP with long-time diffusion (Fig.~\ref{fig:hbabp-ltd}b).
Because the two elements act in series, the particle position separates into two independent contributions, $x(t) = x_{\mathrm{HBABP}}(t) + x_{\mathrm{HW}}(t)$.
Here $x_{\mathrm{HBABP}}(t)$ is the ABP position relative to the HW, and $x_{\mathrm{HW}}(t)$ is the position of the HW center.
Both contributions obey overdamped Langevin equations,
\begin{subequations}
    \begin{align}
        \dot{x}_{\mathrm{HBABP}}(t) & = -\frac{x_{\mathrm{HBABP}}(t)}{\tau_k} + v_{r,\mathrm{HBABP}}(t) + V\cos\phi(t), \label{eq:le-abp-mv-x} \\
        \dot{x}_{\mathrm{HW}}(t)    & = v_{r,\mathrm{HW}}(t). \label{eq:le-abp-mv-hw}
    \end{align}%
    \label{eq:le-abp-mv}%
\end{subequations}%
Here, $V$ is the self-propulsion speed, and $v_{r,\mathrm{HBABP}}(t)$ and $v_{r,\mathrm{HW}}(t)$ are independent white Gaussian velocity noises with zero-mean and correlation $\langle v_{r,i}(t_1)\,v_{r,j}(t_2) \rangle = 2D_i\,\delta_{ij}\,\delta(t_1 - t_2)$, where $D_i$ is the translational diffusivity and $i,j \in \{\mathrm{HBABP}, \mathrm{HW}\}$.
The self-propulsion direction $\hat{V}$ makes an angle $\phi(t)$ with the $x$-axis, and this angle evolves by orientational diffusion, $\dot{\phi}(t) = v_{r,\phi}(t)$, where $v_{r,\phi}(t)$ is a zero-mean white Gaussian noise with $\langle v_{r,\phi}(t_1)\, v_{r,\phi}(t_2) \rangle = 2D_{\mathrm{R}}\,\delta(t_1-t_2)$ and $D_{\mathrm{R}}$ the orientational diffusivity.
The particle equilibrates in the HW over $\tau_k = \gamma_{\mathrm{s}}/k$, the propulsion direction decorrelates over $\tau_{\mathrm{R}} = 1/D_{\mathrm{R}}$, and the medium relaxes over $\lambda = \gamma_{\mathrm{HW}}/k$.
With $a$ the Stokes radius of the ABP, the stiffness follows from the plateau modulus $G_{\mathrm{p}}$, $k = 6\pi a G_{\mathrm{p}}$.
Both frictions obey Stokes relations: the high-frequency friction $\gamma_{\mathrm{s}} = 6\pi a \eta_{\mathrm{s}}$ follows from the solvent viscosity $\eta_{\mathrm{s}}$, and the low-frequency friction $\gamma_{\mathrm{HW}} = 6\pi a \eta_{\mathrm{HW}}$ from the HW viscosity $\eta_{\mathrm{HW}}$, identified with the Maxwell friction $\gamma$.

The MSD of $x_{\mathrm{HBABP}}(t)$ is known for a static HW~\cite{halder2025interplay}, and $x_{\mathrm{HW}}(t)$ contributes free Brownian motion.
Being uncorrelated, the two MSDs add, and for $\tau_{\mathrm{R}} \neq \tau_k$ the 2D MSD follows as
\begin{align}
    \langle \Delta r^2 (\tau)\rangle = {} & 4 D_{\mathrm{HBABP}}\tau \left[1 - e^{-\tau/\tau_k}\right] + \frac{2 V^2 \tau_{k}^2 \tau_{\mathrm{R}}}{\tau_{\mathrm{R}} + \tau_{k}} \times \nonumber             \\
                                          & \left[ 1 - \frac{\tau_{\mathrm{R}} e^{- \tau / \tau_{\mathrm{R}}} - \tau_{k} e^{- \tau / \tau_{k}}}{\tau_{\mathrm{R}} - \tau_{k}} \right] + 4D_{\mathrm{HW}}\tau.
    \label{eq:msd-abp-mv}%
\end{align}%
The first term captures confined thermal motion, the second self-propulsion, and the third long-time diffusion of the HW.
The 2D MSD for the degenerate case $\tau_{\mathrm{R}} = \tau_k \equiv \tau_{\mathrm{d}}$ is derived as
\begin{align}
    \langle \Delta r^2(\tau)\rangle = & 4 D_{\mathrm{HBABP}} \tau \left(1- e^{- \tau/ \tau_{\mathrm{d}}} \right) + V^2 \tau^2_{\mathrm{d}} \times \nonumber \\
                                      & \quad \left[ 1- e^{- t/ \tau_{\mathrm{d}}}\left(1 - \tau/\tau_{\mathrm{d}} \right) \right] + 4D_{\mathrm{HW}}\tau.
    \label{eq:msd-abp-mv-equal}%
\end{align}

We integrate the dynamics with a propagator scheme, advancing the orientation, the diffusing HW, and the relative coordinate at each step.
Setting $V=0$ in Eq.~\eqref{eq:le-abp-mv-x} recovers the HBBP dynamics, whose propagator is the Green's function~\cite{doiTheoryPolymerDynamics2013,chandrasekharStochasticProblemsPhysics1943,halderRealizingMicrorheologicalResponse2026a}
\begin{equation}
    G(x_{\mathrm{HBBP}},t; x_{\mathrm{HBBP}, 0}) =  \left[ 2\pi B(t) \right]^{-1/2} e^{-\frac{(x_{\mathrm{HBBP}} - A(t))^2}{2B(t)}}
    \label{eq:greens-function}
\end{equation}%
with mean $A(t) = x_{\mathrm{HBBP},0}\,e^{-t/\tau_k}$ and variance $B(t) =  [k_{\mathrm{B}}T/k] \left(1-e^{-2t/\tau_k}\right)$.
We propagate $x_{\mathrm{HBABP}}$ with this Green's function, setting $x_{\mathrm{HBBP},0}$ each step from the previous relative position shifted by the active displacement and the HW step.
The relative coordinate advances as $x_{\mathrm{HBABP}}(t_i) = \left[x_{\mathrm{HBABP}}(t_{i-1}) + V\cos\phi(t_{i})\,\Delta t - \Delta x_{\mathrm{HW},i}\right]e^{-\Delta t/\tau_k} + \sqrt{B(\Delta t)}\,R_{\mathrm{HBABP},i}$, where $\Delta x_{\mathrm{HW},i} = \sqrt{2D_{\mathrm{HW}}\,\Delta t}\,R_{\mathrm{HW},i}$ is the HW displacement over the step and the orientation diffuses as $\phi(t_i) = \phi(t_{i-1}) + \sqrt{2D_{\mathrm{R}}\,\Delta t}\,R_{\phi,i}$.
The position of HW center is the cumulative sum $x_{\mathrm{HW}}(t_i) = x_{\mathrm{HW}}(t_{i-1}) + \Delta x_{\mathrm{HW},i}$, and the lab-frame position of the particle is $x(t_i) = x_{\mathrm{HBABP}}(t_i) + x_{\mathrm{HW}}(t_i)$.
Here $R_{\mathrm{HBABP},i}$, $R_{\phi,i}$, and $R_{\mathrm{HW},i}$ are independent Gaussian random numbers with zero-mean and unit variance.
Similarly, lab-frame position of the particle along $y$-axis, i.e, $y(t_i)=y_{\mathrm{HBABP}}(t_i) + y_{\mathrm{HW}}(t_i)$ is calculated by taking the active displacements as $V\sin\phi(t_i)\Delta t$.

\begin{figure*}[t]
    \centering
    \includegraphics[width=\linewidth]{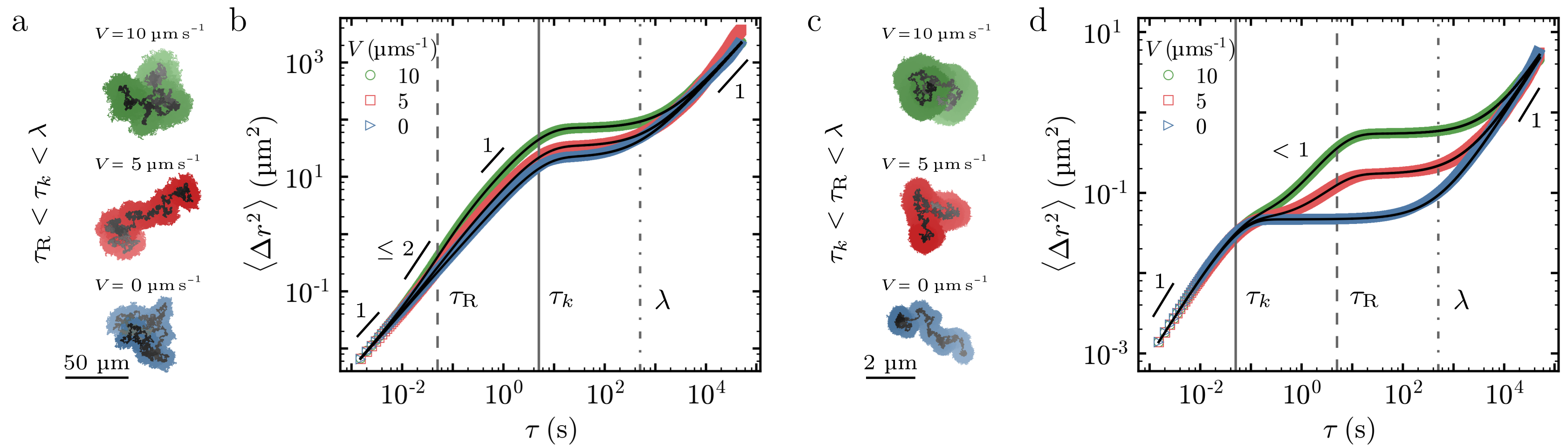}
    \caption{Numerically simulated trajectories and mean square displacements (MSDs) of an HBABP with long-time diffusion for two experimentally accessible timescale orderings: $\tau_{\mathrm{R}} < \tau_k < \lambda$ and $\tau_k < \tau_{\mathrm{R}} < \lambda$.
        In both configurations, MSDs are shown for propulsion speeds $V = \{ 10, 5, 0\}$ $\unit{\micro\meter\per\second}$ at same solvent viscosity $\eta_{\mathrm{s}}=\SI{E-3}{\pascal\second}$.
        At $V = 0$ the MSD reduces to that of a harmonically bound Brownian particle (HBBP) with long-time diffusion.
        Open symbols denote numerical data and solid lines denote fits from Eq.~\ref{eq:msd-abp-mv}.
        Vertical lines mark $\tau_k$ (solid), $\tau_{\mathrm{R}}$ (dashed), and $\lambda$ (dash-dotted).
        (a,~b)~Persistence-dominated regime, $\tau_{\mathrm{R}} < \tau_k < \lambda$ ($\tau_{\mathrm{R}}=\SI{0.05}{\second}$, $\tau_k=\SI{5}{\second}$, $\lambda=\SI{500}{\second}$, $\eta_{\mathrm{HW}}=\SI{0.1}{\pascal\second}$).
        (a)~Simulated trajectories at propulsion speeds $V = 10$ (green), 5 (red), and \SI{0}{\micro\meter\per\second} (blue), with the slowly diffusing HW trajectory overlaid in gray.
        Gradients darken with increasing time.
        (b)~The MSD exhibits successive diffusive, superdiffusive, enhanced-diffusive, plateau, and long-time diffusive regimes separated by the three timescales.
        (c,~d)~Confinement-dominated regime, $\tau_k < \tau_{\mathrm{R}} < \lambda$ ($\tau_k=\SI{0.05}{\second}$, $\tau_{\mathrm{R}}=\SI{5}{\second}$, $\lambda=\SI{500}{\second}$, $\eta_{\mathrm{HW}}=\SI{10}{\pascal\second}$).
        (c)~Same as~(a).
        (d)~The MSD shows a first plateau near $\tau_k$ followed by a second plateau near $\tau_{\mathrm{R}}$, with the plateau values increasing with $V$.
        Long-time diffusive behavior resumes after $\lambda$.}%
    \label{fig:abp-in-mv-sim1}%
\end{figure*}

We simulate the particle dynamics for various configurations of the three timescales $\tau_{\mathrm{R}}$, $\tau_k$, and $\lambda$, at three self-propulsion speeds $V = \lbrace 10, 5, 0 \rbrace$ to $\unit{\micro\meter\per\second}$.
We set each timescale through a distinct physical parameter at fixed temperature $T=\SI{300}{\kelvin}$ and solvent viscosity $\eta_{\mathrm{s}} = \SI{E-3}{\pascal\second}$: the Stokes radius $a$ sets $\tau_{\mathrm{R}}$, the plateau modulus $G_{\mathrm{p}}$ sets $\tau_k = \eta_{\mathrm{s}}/G_{\mathrm{p}}$, and the low-frequency viscosity $\eta_{\mathrm{HW}}$ sets $\lambda = \eta_{\mathrm{HW}}/G_{\mathrm{p}}$.
A fast Fourier transform method yields the MSDs from the simulated trajectories, which accelerates the averaging over more than $\num{e9}$ time-series data points~\cite{calandriniNMoldynInterfacingSpectroscopic2011}.
In the passive limit ($V = 0$), the model reduces to the HBBP with long-time diffusion, and the MSD recovers the known result for a PBP in a single-relaxation VE fluid~\cite{khanRandomWalksColloidal2014,halderRealizingMicrorheologicalResponse2026a}.
In the infinite-relaxation limit ($\lambda \to \infty$), the long-time diffusion vanishes and the model reduces to an HBABP with static HW, recovering the two-timescale interplay of $\tau_k$ and $\tau_{\mathrm{R}}$~\cite{halder2025interplay,halderIdentifyingSignaturesResidual2026}.
These two limits are special cases of the full three-timescale competition, whose regimes we organize by their relative ordering.

\paragraph*{Dynamical regimes.} The dynamics of an HBABP with long-time diffusion are governed by the ordering of three timescales: the equilibration time $\tau_k$, the persistence time $\tau_{\mathrm{R}}$, and the VE relaxation time $\lambda$.
Two orderings are accessible in our experiment, and we devote a subsection to each: persistence-dominated dynamics ($\tau_{\mathrm{R}} < \tau_k < \lambda$) and confinement-dominated dynamics ($\tau_k < \tau_{\mathrm{R}} < \lambda$).
We then treat three additional orderings that complete the physically relevant parameter space.

We first examine persistent dynamics in a weakly restoring dynamic cage, where $\tau_{\mathrm{R}}<\tau_k<\lambda$.
The orientational direction decorrelates at $\tau_{\mathrm{R}}$ before the particle equilibrates at $\tau_k$ in the HW (Fig.~\ref{fig:abp-in-mv-sim1}a, Movie~S1a).
At the shortest timelags the MSD grows diffusively, set by the short-time diffusion coefficient $D_{\mathrm{HBABP}}$.
As active propulsion accumulates over $\tau < \tau_{\mathrm{R}}$, the MSD turns superdiffusive.
Once the orientational direction randomizes near $\tau_{\mathrm{R}}$, dynamics resumes with activity-enhanced diffusion.
This enhanced diffusion persists over $\tau_{\mathrm{R}} < \tau < \tau_k$, while the particle has yet to explore the full extent of the HW.
Near $\tau_k$ the restoring force confines the particle and the MSD develops a plateau. 
This plateau is the VE analog of the elastic confinement observed in passive microrheology~\cite{vanzantenBrownianMotionSingle2000,raikherTheoryBrownianMotion2010,raikherBrownianMotionViscoelastic2013}.
It remains until $\lambda$, beyond which the diffusing HW center restores long-time diffusive growth at a rate set by $D_{\mathrm{HW}}$.
The analytical MSD (Eq.~\ref{eq:msd-abp-mv}, solid lines) captures all regimes quantitatively (Fig.~\ref{fig:abp-in-mv-sim1}b).

Reversing the two timescales $\tau_{\mathrm{R}}$ and $\tau_k$, i.e. $\tau_k<\tau_{\mathrm{R}}<\lambda$, gives confined dynamics in a strongly restoring dynamic cage.
Because the particle encounters the restoring force before its direction randomizes at $\tau_{\mathrm{R}}$, the active force drives it persistently against the restoring potential (Fig.~\ref{fig:abp-in-mv-sim1}c, Movie~S1b)~\cite{halder2025interplay}.
The MSD is diffusive at short timelags and reaches first plateau near $\tau_k$ whose value differs from the passive case.
A second plateau emerges near $\tau_{\mathrm{R}}$ as the orientation fully randomizes within the HW.
After $\lambda$ the diffusing HW center again restores long-time diffusive growth.
The analytical expression (Eq.~\ref{eq:msd-abp-mv}) agrees quantitatively with the numerical data (Fig.~\ref{fig:abp-in-mv-sim1}d).

The contrast between the above two configurations demonstrates that the relative ordering of $\tau_k$ and $\tau_{\mathrm{R}}$ determines whether activity or confinement shapes the intermediate dynamics.
In both cases, $\lambda$ controls the plateau duration and the onset of long-time diffusion, a feature absent in the static-HW limit~\cite{halder2025interplay,halderIdentifyingSignaturesResidual2026}.
We now turn to the remaining three orderings of the timescales and the resulting regimes.

\begin{figure*}[t]
    \centering
    \includegraphics[width=\linewidth]{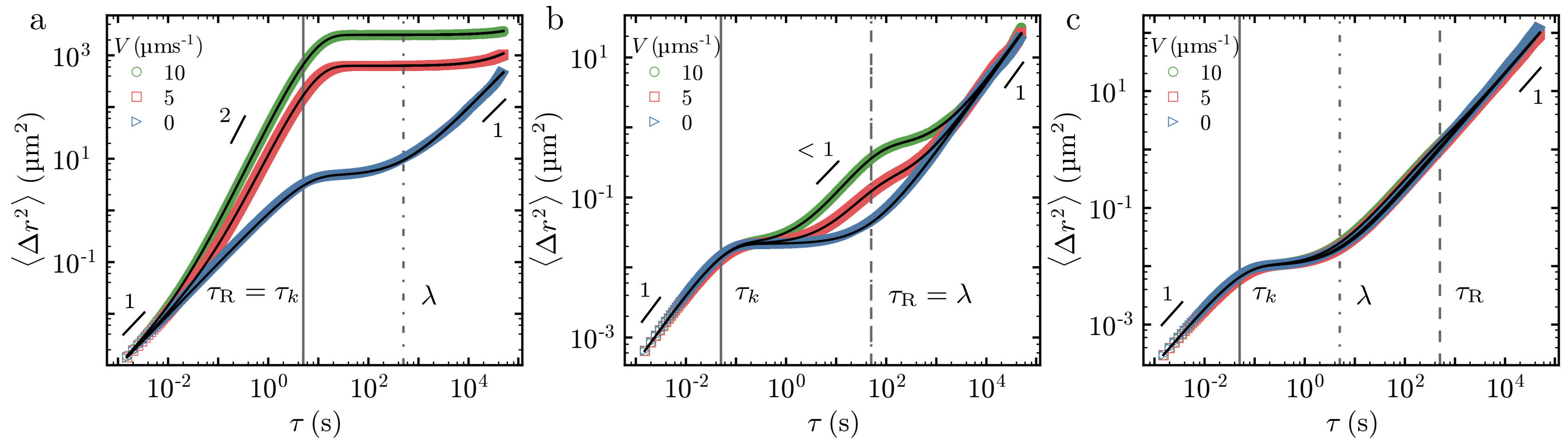}%
    \caption{Numerically simulated MSDs of the HBABP with long-time diffusion for three additional timescale orderings.
        MSDs are shown for $V=\{10, 5, 0\}\;\unit{\micro\meter\per\second}$ at same $\eta_{\mathrm{s}}=\SI{E-3}{\pascal\second}$.
        Open symbols denote numerical data, and solid lines are fits from Eq.~\ref{eq:msd-abp-mv-equal} in~(a) and Eq.~\ref{eq:msd-abp-mv} in~(b,~c).
        (a)~$\tau_{\mathrm{R}} = \tau_k < \lambda$ ($\tau_{\mathrm{R}}=\tau_k=\SI{5}{\second}$, $\lambda=\SI{500}{\second}$, $\eta_{\mathrm{HW}}=\SI{0.1}{\pascal\second}$).
        The MSD shows diffusive, ballistic, plateau, and long-time diffusive regimes, all shifting to larger MSD values with increasing $V$.
        (b)~$\tau_k < \tau_{\mathrm{R}} = \lambda$ ($\tau_k=\SI{0.05}{\second}$, $\tau_{\mathrm{R}}=\lambda=\SI{50}{\second}$, $\eta_{\mathrm{HW}}=\SI{1}{\pascal\second}$).
        The MSD develops two plateaus at $\tau_k$ and $\lambda$ with increasing $V$, and the crossover to long-time diffusion shifts beyond $\lambda$.
        (c)~$\tau_k < \lambda < \tau_{\mathrm{R}}$ ($\tau_k=\SI{0.05}{\second}$, $\lambda=\SI{5}{\second}$, $\tau_{\mathrm{R}}=\SI{500}{\second}$, $\eta_{\mathrm{HW}}=\SI{0.1}{\pascal\second}$).
        The MSDs for different $V$ collapse onto the passive curve, showing no prominent activity signature.}%
    \label{fig:abp-in-mv-sim2}%
\end{figure*}

When $\tau_{\mathrm{R}} = \tau_k  < \lambda$, equilibration and reorientation of the particle occur simultaneously (Movie~S1c).
The MSD progresses through short-time diffusion, an intermediate ballistic regime, and a plateau, before crossing over to long-time diffusion beyond $\lambda$ (Fig.~\ref{fig:abp-in-mv-sim2}a).
Increasing the propulsion speed $V$ shifts these features to larger MSD values because the activity-enhanced diffusivity at the plateau grows with $V$ and exceeds $D_{\mathrm{HW}}$ at $\lambda$, delaying the crossover to long-time diffusion until the two contributions become comparable.
The analytical MSD (Eq.~\ref{eq:msd-abp-mv-equal}, solid lines) captures this behavior.

For $\tau_{\mathrm{R}} = \lambda$ (with $\tau_k < \tau_{\mathrm{R}}$), particle orientation randomization and VE relaxation coincide (Movie~S1d).
With increasing $V$, the MSD evolves from a plateau at $\tau_k$ to another plateau at $\tau_{\mathrm{R}}=\lambda$, and the onset of long-time diffusion shifts beyond $\lambda$ (Fig.~\ref{fig:abp-in-mv-sim2}b).
Because orientation randomization and long-time relaxation occur together, the effective diffusivity at activity-enhanced plateau region shifts to long-time diffusivity $D_{\mathrm{HW}}$, which becomes pronounced at large $V$.

In the last ordering, $\lambda < \tau_{\mathrm{R}}$ (with $\tau_k < \lambda$), the relaxation happens before the particle's reorientation (Movie~S1e).
The long persistence time means the particle has not yet randomized its orientation when the HW begins to diffuse at $\lambda$ (Fig.~\ref{fig:abp-in-mv-sim2}c).
As a result, the activity-enhanced diffusivity at intermediate timelags remains comparable to $D_{\mathrm{HW}}$, and no prominent superdiffusive signature appears in the MSD.
The dynamics effectively reduce to those of an HBBP with long-time diffusion.

\paragraph*{Active Janus colloid in configurable VE medium.} A thermophoretically active particle is realized by a metal-coated Janus colloid in an optical trap~\cite{jiangActiveMotionJanus2010}.
However, its stable trapping is challenging because the two hemispheres interact differently with the trapping laser~\cite{brontecirizaOpticallyDrivenJanus2023,halderDynamicallyStableOptical2026}.
We use Pt-silica Janus colloids ($2a \sim \SI{1.76}{\micro\meter}$) with a $\approx \SI{5.5}{\nano\meter}$ thick Pt cap on one hemisphere.
The Pt coating absorbs and reflects more radiation than the bare silica~\cite{wernerOpticalConstantsInelastic2009,kulikovaOpticalPropertiesTungsten2020}, generating a temperature gradient across the particle that drives thermophoretic propulsion within the trap itself~\cite{jiangActiveMotionJanus2010}.
Adding \ce{H2O2} enhances the propulsion speed through diffusiophoresis driven by the solute concentration gradient around the particle~\cite{howseSelfMotileColloidalParticles2007}.

Stable trapping requires a balance among the optical forces, namely the gradient and scattering forces, and the thermophoretic force~\cite{halderDynamicallyStableOptical2026}.
Their interplay keeps the particle localized about the trap center at low laser power $P$ and pushes it to a force-balanced radial distance at higher $P$, producing a delocalized confinement~\cite{halderDynamicallyStableOptical2026}.
Boltzmann inversion of the measured position distributions yields an effective harmonic potential in both cases, validating the HBABP framework~\cite{halder2025interplay}.
Increasing $P$ stiffens the trap, shortening $\tau_k = \gamma_{\mathrm{s}}/k$, and enhances the thermophoretic propulsion speed $V$, which increases further with \ce{H2O2} concentration through diffusiophoresis.
$\tau_{\mathrm{R}}$ depends mainly on the particle size and the nonuniform heating of the surrounding fluid by the Pt cap.
The resulting orderings of $\tau_{\mathrm{R}}$ and $\tau_k$ select two dynamical states~\cite{halder2025interplay}.
For $\tau_{\mathrm{R}} < \tau_k$, the particle isotropically fills the confinement about the trap center, and the position distribution is Gaussian.
For $\tau_{\mathrm{R}} > \tau_k$, it stays on the annulus at the force-balanced radial distance, and the distribution is bimodal.

\begin{figure*}[t]
    \centering
    \includegraphics[width=\linewidth]{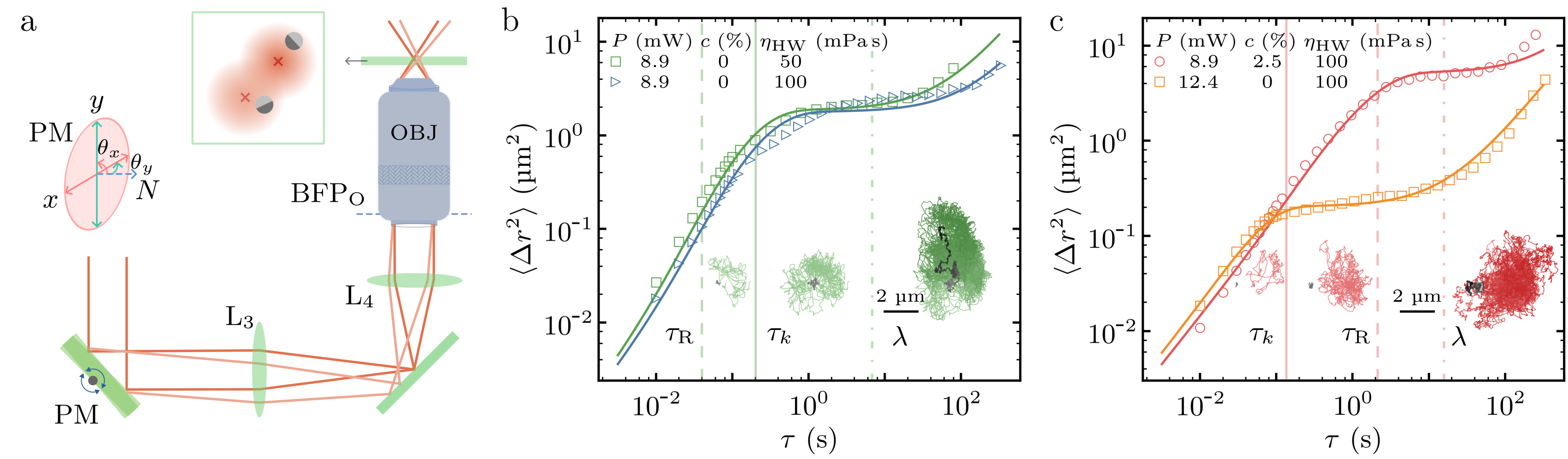}%
    \caption{Experimental realization of a phoretically active Pt-coated silica Janus colloid in a configurable single-relaxation VE medium.
        (a)~Schematic of the setup: a piezo-controlled mirror (PM) is conjugated to the back-focal plane of the objective ($\mathrm{BFP}_{\mathrm{O}}$) through the afocal system of lenses $\mathrm{L}_3$ and $\mathrm{L}_4$, and its angular displacement (left inset) steers the trap center along a computer-generated Brownian trajectory at the sample plane (top inset), producing a dynamic optical trap with friction coefficient $\gamma_{\mathrm{HW}}$.
        (b,~c)~MSDs of a Pt-coated Janus colloid in the dynamic trap, with trajectory insets.
        Open symbols denote experimental data and solid lines fits to Eq.~\ref{eq:msd-abp-mv}.
        Vertical lines mark $\tau_k$ (solid), $\tau_{\mathrm{R}}$ (dashed), and $\lambda$ (dash-dotted).
        (b)~Persistence-dominated regime, $\tau_{\mathrm{R}} < \tau_k < \lambda$.
        Inset: particle trajectory (green) and trap center trajectory (gray) over successive durations $t = \lbrace 1, 10, 100 \rbrace \; \unit{\second}$, shown light to dark with increasing $t$, at laser power $P=\SI{8.9}{\milli\watt}$.
        The MSD exhibits successive diffusive, superdiffusive, enhanced-diffusive, plateau, and long-time diffusive stages, matching the numerical prediction of Fig.~\ref{fig:abp-in-mv-sim1}b.
        (c)~Confinement-dominated regime, $\tau_k < \tau_{\mathrm{R}} < \lambda$.
        The plateau at $\tau_k$ is obscured by the large $V$ and nearby $\tau_{\mathrm{R}}$, and the MSD shows diffusive recovery beyond $\lambda$, matching the numerical prediction of Fig.~\ref{fig:abp-in-mv-sim1}d.
        Inset: particle trajectory (red) and trap center trajectory (gray) as in (b), at $P=\SI{8.9}{\milli\watt}$ and \ce{H2O2} concentration $c=2.5\%$.}%
    \label{fig:abp-in-mv-expt}%
\end{figure*}

The HBABP with long-time diffusion requires the HW itself to diffuse slowly, realizing the Maxwell relaxation of the MV fluid~\cite{halderRealizingMicrorheologicalResponse2026a}.
We achieve this with a dynamic optical trap, feeding a computer-generated Brownian trajectory as angular displacements to a piezo-controlled mirror (PM) conjugated to the back-focal plane of the objective ($\mathrm{BFP}_{\mathrm{O}}$), which produces proportional lateral displacements of the trap center  (Fig.~\ref{fig:abp-in-mv-expt}a)~\cite{fallmanDesignFullySteerable1997, halderOpticalMicromanipulationSoft2024a, halderRealizingMicrorheologicalResponse2026a}.
The trap center consequently executes Brownian motion at the sample plane, with an effective friction coefficient $\gamma_{\mathrm{HW}}$ set by the generated trajectory.
A rapidly moving trap cannot hold the self-propelled particle, so we choose $\gamma_{\mathrm{HW}}$ large enough that the particle stays confined and $\lambda$ remains the longest timescale.
Particle positions are tracked from recorded images using the TrackMate plugin in Fiji~\cite{ershovTrackMate7Integrating2022}.
We compute the time-averaged MSD from the tracked trajectories and fit it to Eq.~\ref{eq:msd-abp-mv}, obtaining $\tau_k$, $\tau_{\mathrm{R}}$, $\lambda$, and $V$ with fitting uncertainties below 10\%.
Tuning $P$ and $\gamma_{\mathrm{HW}}$ realizes both orderings, $\tau_{\mathrm{R}} < \tau_k < \lambda$ and $\tau_k < \tau_{\mathrm{R}} < \lambda$.

In the persistence-dominated regime, $\tau_{\mathrm{R}} < \tau_k < \lambda$, the persistent motion of the trajectory isotropically fills the slowly diffusing confinement (Fig.~\ref{fig:abp-in-mv-expt}b, inset; Movie~S2a).
All numerically identified dynamical stages reappear in the experimental MSD: short-time diffusion, superdiffusion, enhanced diffusion, the confinement plateau, and long-time diffusive recovery (Fig.~\ref{fig:abp-in-mv-expt}b).
Weak trap stiffness places $\tau_k$ well above $\tau_{\mathrm{R}}$, so the particle randomizes its orientation before it encounters the restoring force.
Consequently, the full active signature develops ahead of the plateau.
Both the analytical MSD (Eq.~\ref{eq:msd-abp-mv}) and the corresponding numerical result (Fig.~\ref{fig:abp-in-mv-sim1}b) agree quantitatively with the experimental data.

In the confinement-dominated regime, $\tau_k < \tau_{\mathrm{R}} < \lambda$, the restoring force confines the particle before its orientation randomizes.
Accordingly, the trajectory traces short arcs of the annulus, which the trap diffusion displaces on the scale of $\lambda$ (Fig.~\ref{fig:abp-in-mv-expt}c, inset; Movie~S2b).
The MSD shows short-time diffusion, a plateau at $\tau_{\mathrm{R}}$, and diffusive recovery beyond $\lambda$ (Fig.~\ref{fig:abp-in-mv-expt}c).
Higher $V$ and closely spaced $\tau_k$ and $\tau_{\mathrm{R}}$ leave the first plateau, at $\tau_k$, unresolved.
Analytical (Eq.~\ref{eq:msd-abp-mv}) and numerical predictions (Fig.~\ref{fig:abp-in-mv-sim1}d) again describe the experimental MSD quantitatively.

The experimentally emulated dynamics thus reproduce the simulated ABP behavior in a single-relaxation VE fluid, validating the HBABP with long-time diffusion model without any real VE medium.

\paragraph*{Conclusions.}
We studied the dynamics of an ABP in a single-relaxation VE fluid, combining analytical theory, numerical simulation, and experiment on a Pt-coated Janus colloid in a dynamic optical trap.
The competition among three timescales governs these dynamics: the equilibration time $\tau_k$, the persistence time $\tau_{\mathrm{R}}$, and the VE relaxation time $\lambda$.
Their relative ordering selects qualitatively distinct intermediate regimes, identifiable in the MSD.
The HBABP model with long-time diffusion captures all simulated regimes quantitatively.
Experiment tests the case $\lambda > \tau_k, \tau_{\mathrm{R}}$, with $\lambda$ known from the trap rather than fitted.

The present study is restricted to linear viscoelasticity with a single relaxation time~\cite{khanRandomWalksColloidal2014, halderRealizingMicrorheologicalResponse2026a}.
A single dynamic trap cannot reproduce nonlinear responses such as shear-thinning or strain-stiffening~\cite{khanOpticalTweezersMicrorheology2019, dattSquirmingShearthinningFluids2015}, and the laser power couples the trap stiffness to the propulsion speed~\cite{halder2025interplay}.
The analysis assumes the dilute, single-particle limit in the plane, with no interparticle or hydrodynamic coupling.
Natural extensions include generalized linear VE media with multiple relaxation modes~\cite{khanTrajectoriesProbeSpheres2014}, the collective dynamics of many active particles~\cite{gompper2020MotileActive2020}, microrheology~\cite{squiresFluidMechanicsMicrorheology2010, furstMicrorheology2020} with self-propulsion replacing external forcing, and the energy exchange between probe and environment~\cite{schuttlerActiveParticlesMoving2025, darabiStochasticEnergeticsColloidal2023}.
Wherever self-propulsion meets fluid memory, these three timescales and their ordering select the dynamical regime.

\paragraph*{Acknowledgments.} The authors acknowledge the SERB (now ANRF), Govt.\ of India, for supporting this work through a Core Research Grant (Grant No.\ CRG/2020/002723), and the PARAM Sanganak computing facility at the Computer Center, IIT Kanpur, for the numerical simulations.
M.K. acknowledges funding from IIT Kanpur through an initiation grant (Grant No.\ IITK-PHY-2017081).

All authors contributed to the conception and design of the research.
S.H. performed the numerical simulations and experiments, and analyzed the data.
S.H. and M.K. interpreted the data and wrote the manuscript.
M.K. supervised the project.

\paragraph*{Data availability.} The data that support the findings of this study are not publicly available because of the large size of the raw microscopy images and simulation data. They are available from the authors upon reasonable request.

\bibliography{refs.bib}

\end{document}